\documentclass[aps,pre,showpacs,noshowkeys,amsmath,amssymb,amsfonts,superscriptaddress,longbibliography,reprint]{revtex4-2}
\usepackage[english]{babel}
\UseRawInputEncoding
\usepackage{graphicx}
\usepackage{bm}
\usepackage{physics}
\usepackage{siunitx}
\usepackage{mathtools}
\usepackage{gensymb}

\newcommand{\Ta}{\ensuremath{T_\text{a}}}
\newcommand{\Vs}{\ensuremath{V_\text{S}}}
\newcommand{\Vt}{\ensuremath{V_\text{T}}}
\newcommand{\Vavg}{\ensuremath{\overline{V}}}

\newcommand{\Ca}{\ensuremath{\text{Ca}}}
\newcommand{\hc}{\ensuremath{h_\text{c}}}

\def\showcom{} 

\ifdefined\showcom
  \newcommand{\comr}[1]{\textcolor{purple}{#1}}
  \newcommand{\comb}[1]{\textcolor{blue}{#1}}
\else
  \newcommand{\comr}[1]{}
  \newcommand{\comb}[1]{}
\fi

\usepackage{etoolbox}
\AtBeginDocument{%
}
\usepackage{caption}
\usepackage{subcaption}
\DeclareCaptionLabelSeparator{bar}{~\rule[-0.4ex]{0.2ex}{1em}~}
\DeclareCaptionLabelFormat{subfor}{\textbf{#2}}
\newcommand*\bfcaption[2]{\caption[#1]{\textbf{#1.}#2}}
\usepackage{xcolor}
\definecolor{UBcolor}{HTML}{007CC1}

\usepackage[colorlinks=true,pdfnewwindow=true,linkcolor=UBcolor,citecolor=UBcolor,urlcolor=UBcolor,breaklinks=true,linktocpage]{hyperref}
\usepackage[all]{hypcap}
\usepackage[nameinlink,capitalise]{cleveref}
\AddToHook{cmd/appendix/before}{%
  \crefalias{section}{appendix}%
  \crefalias{subsection}{appendix}%
}
\usepackage{xr}
\begin{document}

\title{Migrating cell clusters as active droplets with asymmetric shape and contact angles}

\author{Hanno I. Hennighausen}
\altaffiliation{These authors contributed equally to this work.}
\affiliation{Max Planck Institute for the Physics of Complex Systems, N\"{o}thnitzerst. 38, 01187 Dresden, Germany}
\affiliation{Center for Systems Biology Dresden, Pfotenhauerst. 108, 01307 Dresden, Germany}

\author{Jaime A. Espina}
\altaffiliation{These authors contributed equally to this work.}
\affiliation{Cluster of Excellence Physics of Life, Technische Universit\"{a}t Dresden, 01062 Dresden, Germany}

\author{Elias H. Barriga}
\email{elias.barriga@tu-dresden.de}
\affiliation{Cluster of Excellence Physics of Life, Technische Universit\"{a}t Dresden, 01062 Dresden, Germany}

\author{Ricard Alert}
\email{ricard.alert@ub.edu}
\affiliation{Max Planck Institute for the Physics of Complex Systems, N\"{o}thnitzerst. 38, 01187 Dresden, Germany}
\affiliation{Center for Systems Biology Dresden, Pfotenhauerst. 108, 01307 Dresden, Germany}
\affiliation{Cluster of Excellence Physics of Life, Technische Universit\"{a}t Dresden, 01062 Dresden, Germany}
\affiliation{Departament de F\'{i}sica de la Mat\`{e}ria Condensada, Universitat de Barcelona, Barcelona, Spain}
\affiliation{Universitat de Barcelona Institute of Complex Systems (UBICS), Barcelona, Spain}
\affiliation{Instituci\'{o} Catalana de Recerca i Estudis Avan\c{c}ats (ICREA), Barcelona, Spain}

\date{\today}

\begin{abstract}
During embryonic development and cancer progression, clusters of cells migrate collectively following external cues. Such migrating cell clusters have been modeled as active droplets wetting a substrate. So far, this active wetting model assumed the cluster to take the symmetric shape of a spherical cap. Here, in contrast, we experimentally show that migrating clusters of frog neural crest cells have asymmetric height profiles, with a lower contact angle at the front than at the rear. We then extend the active wetting model by deriving an equation for the cluster shape, which becomes part of the solution rather than an assumption. Our model shows that shape asymmetry can arise from the asymmetric tractions that drive cluster migration: They generate a pressure gradient across the tissue which, via the Young-Laplace equation, leads to a curvature gradient on the cluster's upper surface. Finally, by fitting the theoretical predictions to the measured height profile, we estimate the tissue viscosity. Overall, our results provide a general mechanism for the emergence of asymmetric shapes in migrating cell clusters, understood as active droplets on a wettability gradient.
\end{abstract}

\maketitle

Collective cell migration drives key steps in cancer invasion, tissue regeneration, and development \cite{Friedl2009,Scarpa2016,Mayor2016,Cheung2025,Weissenbruch2025}. Whereas in some cases a cluster of cells spreads in all directions, in other cases the cluster migrates directionally as a whole. Examples include cranial neural crest (NC) cells in mice and frogs, the zebrafish lateral line, and \textit{Drosophila} border cells \cite{Friedl2009,Scarpa2016,Mayor2016,Cheung2025,Weissenbruch2025}. In all these cases, the directional migration of a cell cluster is key for the development of organs and tissues. Directional cluster migration is also studied in vitro, where cells follow chemical, mechanical, and electrical cues by performing chemotaxis \cite{Theveneau2010,Malet-Engra2015,Shellard2018}, durotaxis \cite{Shellard2021,Pallares2023}, and electrotaxis \cite{Zajdel2020,Shim2021}, respectively.

To understand the mechanics of these processes, recent work developed the theoretical framework of active wetting, in which cell aggregates are modelled as active droplets wetting a substrate \cite{Perez-Gonzalez2019,Pallares2023,Franco-Onate2026,Alert2020,Pahlavan2026}. In this model, active traction forces drive collective cell migration against surface tension and contractile cell-cell stresses. This model has been used to explain tissue wetting transitions \cite{Perez-Gonzalez2019,Franco-Onate2026} and epithelial fingering \cite{Perez-Gonzalez2019,Alert2019}, as well as the spreading of cancer cell aggregates \cite{Marchesi2026}, ectoderm tissue \cite{Le2026}, and mouse embryos \cite{Cavanaugh2026}. In this framework, directed cluster migration results from an asymmetry in active tractions between the front and rear of the cluster \cite{Beaune2018,Rossetti2024}. For example, durotaxis results from stronger traction on stiffer regions of the substrate \cite{Alert2019a,Pi-Jauma2022,Pallares2023}. So far, these models assumed cell clusters to be shaped as spherical caps, as for sessile droplets shaped solely by surface tension.

Here, we ask: What is the shape of migrating cell clusters? By examining clusters of neural crest cells performing durotaxis in vitro, we found that their shape departs from that of a spherical cap. The clusters are asymmetric: They are thicker towards the rear and thinner towards the front, leading to a higher contact angle at the rear than at the front. To explain this finding, we extend active wetting theory to account for the flows that the basal layer of motile cells induces on the upper cell layers \cite{Yousafzai2024}. We show that the induced flows create a pressure gradient across the cluster: Pressure is higher at the rear, which results in higher curvature of the cluster's top surface, and hence a higher contact angle at the rear. Building on classic work on dynamic contact angles in wetting \cite{Leger1992}, we derive a nonlinear equation for the cluster shape, whose solution fits our experimental data and allows us to estimate the cluster's viscosity. Overall, our work puts forward a physical picture of migrating cell clusters as active droplets on a wettability gradient: The front-rear difference in active traction that drives cluster migration also induces flows that shape it. These results provide a general mechanism, based on the physics of active wetting, whereby migrating cell clusters acquire their shape.

\bigskip

\label{sec: experiments}

\begin{figure*}[tbp!]
\begin{center}
\includegraphics[width=\textwidth]{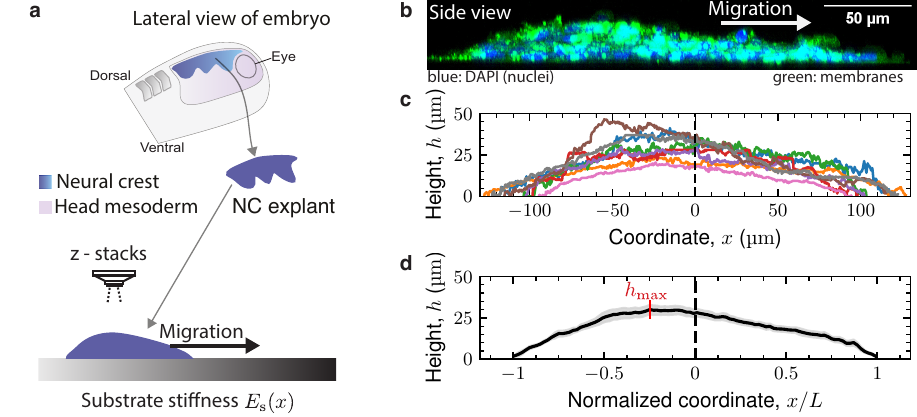}
\end{center}
 {\phantomsubcaption\label{fig: setup}}
 {\phantomsubcaption\label{fig: exp_image}}
 {\phantomsubcaption\label{fig: height_profiles}}
 {\phantomsubcaption\label{fig: height_profiles_avg}}
\bfcaption{Shape asymmetry in migrating cell clusters}{ \subref*{fig: setup}, Experimental setup. 
\subref*{fig: exp_image}, Side view of a migrating NC cell cluster. 
\subref*{fig: height_profiles}, Height profiles of 8 cell clusters. 
\subref*{fig: height_profiles_avg}, Averaged height profile. The gray shade indicates the standard error of the mean.}\label{fig: 1_experimental_height}
\end{figure*}

\noindent\textbf{Asymmetric shape of migrating cell clusters.} We study the shape of migrating clusters of cranial NC cells from the frog \textit{Xenopus laevis}.
The NC is a pluripotent motile cell population that migrates through the embryo during development. It follows multiple signals \cite{Barriga2020,Weissenbruch2025}, including chemotactic \cite{Theveneau2010,Barriga2018,Shellard2021}, durotactic \cite{Shellard2021}, and electrotactic \cite{Ferreira2025} cues. 
Here, we take explants of NC cells in the migratory stage and plate them on polyacrylamide substrates with a stiffness gradient (\cref{fig: setup}, \cref{methods}). 
In these conditions, NC clusters migrate towards the stiffer region of the substrate --- a process called durotaxis \cite{Shellard2021,Espina2022}. 
By fixing the migrating clusters and imaging them with confocal microscopy (\cref{methods}), we reconstructed their three-dimensional shape (\hyperref[movies]{Video 1}). 
The side view revealed that NC clusters have a front-back asymmetric shape (\cref{fig: exp_image}).

To quantify cluster shape, we measured the height profile $h(x)$ of eight clusters along their direction of motion (\cref{fig: height_profiles}). For each cluster, we measured the contact angles and found that, on average, the rear one, $\theta_\text{r} = (24 \pm 3)^\circ$, is slightly larger than the front one, $\theta_\text{f} = (18 \pm 5)^\circ$ (\cref{fig: individual contact angles}). We also quantified the position of the highest point; it is shifted rearwards, with an average of $x_\text{max} = -32\pm 7\, \si{\um}$ (\cref{fig: individual contact angles}). These results show that the height profile is asymmetric.

Since each cluster had a slightly different length, we then averaged the height profiles by first normalizing the spatial coordinate $x$ of each cluster by its half-length $L$ and then binning data points into 100 bins (\cref{fig: height_profiles_avg}). We found a mean cluster half-length of $L = 110 \pm 11 \, \si{\um}$. For the averaged height profile, the peak is also shifted rearwards, at $x_\text{max} = - 28 \pm 2 \, \si{\um}$, with an average height of $h_\text{max} = 30 \pm 3 \, \si{\um}$, and its rear and front contact angles are also slightly different: $\theta_\text{r} = (25 \pm 2)^\circ$ and $\theta_\text{f} = (22 \pm 2)^\circ$ (\cref{fig: contact angles}). Below, we propose a theory to explain this shape asymmetry.

\bigskip

\label{sec: Model}

\begin{figure}[tb]
\begin{center}
\includegraphics[width=\columnwidth]{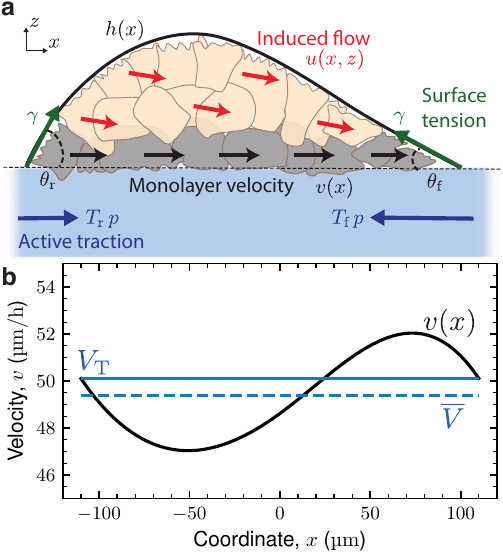}
\end{center}
 {\phantomsubcaption\label{fig: 2_model_schematic}}
 {\phantomsubcaption\label{fig: velocity_field}}
\bfcaption{Active droplet model for migrating cell clusters}{ 
\subref*{fig: 2_model_schematic}, Schematic of the model, which distinguishes the basal cell layer (gray) from the upper cell layers (beige). Cells in the basal layer are in contact with the substrate (light blue), where they exert front-back asymmetric tractions (blue arrows) that drive the monolayer flow (black arrows) that result in cluster migration.
The flow in the basal layer induces flows in the upper layers (red arrows), which ultimately lead to the asymmetric height profile. The upper surface of the cluster has surface tension (green arrows).
\subref*{fig: velocity_field}, Velocity profile (black), translation velocity (blue) and average velocity (dashed blue) of the basal monolayer predicted by our model. Parameter values, listed in \cref{tab: parameter values}, are chosen to match experiments.}\label{fig: model_velocity}
\end{figure}

\noindent\textbf{Active droplet model for migrating cell clusters.} Our approach is to explain the observed shape asymmetry as arising from the asymmetric active tractions that drive cluster migration, which shape the tissue through the physics of wetting. To this end, we extend the theory of active wetting to allow the cell cluster to be shaped by its internal flows. Following previous work \cite{Pallares2023,Yousafzai2024}, we model the cell cluster as consisting of two parts: (i) a basal layer of cells that actively migrate on the substrate, and (ii) upper layers of cells that remain non-migratory (\cref{fig: 2_model_schematic}).

We model the basal layer through the active wetting model proposed in previous works \cite{Perez-Gonzalez2019,Pallares2023,Cavanaugh2026,Franco-Onate2026}. Force balance for the basal cell layer in the direction of migration reads
\begin{equation}\label{eq: 1d force balance}
    \partial_x \sigma = \xi v(x) - \Ta(x) p(x),
\end{equation}
where $v(x)$ is the monolayer flow field. Here, the internal forces due to viscous tension $\sigma = \eta \partial_x v$, with monolayer viscosity $\eta$, balance the forces exchanged between the cells and the substrate, which include friction with a coefficient $\xi$, and active tractions of magnitude $\Ta(x)$ exerted by cells with front-back polarity as described by the field $p(x)$. Compared to previous work \cite{Perez-Gonzalez2019,Franco-Onate2026}, here we ignore contractile cell-cell active stresses, since we do not need them to explain shape asymmetry. As boundary conditions for the monolayer tension $\sigma$, we impose a generalized Young-Dupr\'{e} condition at the front and rear:
\begin{subequations} \label{eq Young-Dupre}
\begin{align}
    \sigma(x=+ L) =& -\gamma \cos{\theta_\text{f}},\\
    \sigma(x=- L) =& -\gamma \cos{\theta_\text{r}}.
\end{align}
\end{subequations}
Here, $\gamma$ is the tissue surface tension, and we have allowed the front and rear contact angles, $\theta_\text{f}$ and $\theta_\text{r}$, to be different (\cref{fig: 2_model_schematic}).

Regarding active tractions, we take a polarity field $p(x) = x/L$, so that cells at the monolayer edges are maximally polarized outwards, $p(x=\pm L) = \pm 1$. Compared to previous work, the linear behavior chosen here corresponds to the limit of large polarity decay length \cite{Perez-Gonzalez2019}, which is sufficient for our purposes. Given that the polarity field is symmetric, net cluster migration requires an asymmetry in traction magnitude $\Ta(x)$ \cite{Rossetti2024}. Previous works modeled durotaxis by taking the active traction magnitude to increase with substrate stiffness $E_\text{s}$ \cite{Alert2019a,Pi-Jauma2022,Pallares2023}. Here, to simplify the model, we take a piece-wise profile in which the front half of the cluster exerts stronger traction than the rear half:
\begin{equation} \label{eq: traction_front_back}
T_\text{a}(x) =
\begin{cases}
T_{\text{r}}, &  x \in [-L,0] \\
T_{\text{f}}, & x \in [0,L].
\end{cases}
\end{equation}
Here, $x$ is the comoving coordinate, so that $x=0$ marks the cluster center at all times.

We solve the model analytically to obtain the velocity profile $v(x)$ of the basal cell layer (\cref{sec: velocity_solution}). Here, we focus on steady translating solutions, for which the cluster migrates while keeping a fixed length $2L$. In this case, the spreading velocity $\Vs = \left[v(L)-v(-L)\right]/2$ vanishes, and the translation velocity is given by that of both front and rear edges: $\Vt = v(L) = v(-L)$. We chose parameter values as explained in \cref{sec: parameter_estimation} to obtain a steady translating solution and match experimental measurements including $L \approx 110$ $\si{\um}$ and $\Vt \approx 50$ $\si{\um}$/h. \Cref{fig: velocity_field} shows the resulting velocity field $v(x)$ at steady state.

\bigskip

\noindent\textbf{Induced flow in the upper cell layers.} The velocity field $v(x)$ of the basal layer, which represents the cluster migration, induces flow in the upper cell layers (\cref{fig: 2_model_schematic}). Following recent work \cite{Yousafzai2024}, we treat these upper layers as a passive incompressible fluid with viscosity $\mu$ at low Reynolds number, so that its flow field $\mathbf{u}$ follows the Stokes equation
\begin{equation} \label{eq: stokes}
    \mu \nabla^2 \mathbf{u} - \nabla P = 0,
\end{equation}
where $P$ is the pressure that enforces the incompressibility condition $\nabla \cdot \mathbf{u} =  0$. Note that the cluster viscosity $\mu$ is different from the monolayer viscosity $\eta$ introduced before, which captures the dissipation associated with the active flows in the basal cell layer, including its expansion and retraction. We consider a vertical cross-section of the cluster so that $\mathbf{u}= (u_x(x,z), u_z(x,z))$, and we impose the following boundary conditions: (i) no vertical flow into or out of the substrate: $u_z(x,0) = 0$; (ii) the horizontal flow matches that of the basal layer at the bottom: $u_x(x,0) = v(x)$; and (iii) shear stresses vanish at the upper surface of the cluster, since it is a free interface: $\sigma_{nt}(x,h(x)) = 0$, where $n$ and $t$ respectively denote the axes normal and tangential to the interface $h(x)$ (\cref{fig: 2_model_schematic}). Below, we work in the lubrication approximation, or thin-film limit \cite{DeGennes1985,Guyon2001}, which is appropriate given the small height-to-length ratio $\epsilon = h_\text{max}/(2L) \approx 0.13 $ of our clusters (\crefrange{fig: exp_image}{fig: height_profiles_avg}). In this limit, condition (iii) reduces to $\partial_z u_x (x,h(x)) = 0$.

\bigskip

\noindent\textbf{The droplet shape equation.} We now use the lubrication approximation to derive an equation for the cluster's shape $h(x)$ given its basal migration flow $v(x)$. To this end, we start with the thin-film equation \cite{Guyon2001},
\begin{equation}
\partial_t h(x,t) = - \partial_x \int_0^{h(x,t)} u_x(x,z) \,\dd z,
\end{equation}
which uses incompressibility to relate changes in the droplet height to the flow. From this equation, one can obtain the Reynolds equation \cite{Guyon2001}, which in our case reads
\begin{equation}\label{eq: reynolds}
    \partial_x \left( h^3 \partial_x P \right) = 3 \mu \left[ \partial_t h + \partial_x (h v)\right].
\end{equation}
Since the droplet migrates at speed $\Vt$, the steady-state height profile obeys $h(x,t) = h(x - \Vt t)$, so that $\partial_t h = -\Vt \partial_x h$. Integrating from $x=-L$, where $h(-L)=0$, to $x$, we obtain
\begin{equation}\label{eq: reynolds_integrated}
    P'(x) = -3 \mu \frac{\Vt-v(x)}{h^2(x)}.
\end{equation}
This equation has two unknowns, $P(x)$ and $h(x)$. However, these two quantities are related via the Young-Laplace equation, $P(x) = \gamma K(x)$, which relates the pressure to the interface curvature $K(x)$ through the surface tension $\gamma$ (\cref{fig: 2_model_schematic}). Approximating for small gradients, we have $K \approx - h''(x)$, and hence $P(x) \approx - \gamma h''(x)$. Introducing this relation in \cref{eq: reynolds_integrated}, we obtain a closed equation for the droplet shape:
\begin{equation} \label{eq: shape}
    h'''(x) =3 \Ca \frac{1- v(x)/\Vt}{h^2(x)}.
\end{equation}
Here, we have defined the capillary number $\Ca =  \mu \Vt/\gamma$, which compares viscous and capillary forces. \Cref{eq: shape} is a variant of the shape equation used to predict dynamic contact angles in spreading droplets \cite{Leger1992}. Unlike in that case, however, we did not impose a no-slip boundary condition at the substrate, and the velocity that enters the equation is not the depth-averaged one but that of the basal cell layer, which drives the rest of the droplet.

\Cref{eq: reynolds_integrated,eq: shape} reveal the mechanism of the shape asymmetry. For a cluster with a uniform basal velocity, $v(x) = \Vt$, the pressure gradient vanishes (\cref{eq: reynolds_integrated}). Hence, the cluster has a uniform curvature ($h'''=0$ in \cref{eq: shape}) and is therefore symmetric. In contrast, any non-homogeneous basal velocity field $v(x)$ gives rise to a pressure gradient (\cref{eq: reynolds_integrated}). Through the Young-Laplace equation, $P(x) \approx -\gamma h''(x)$, a pressure gradient gives rise to a curvature gradient at the droplet's interface, which implies $h'''\neq 0$ in \cref{eq: shape} and hence an asymmetric shape. Thus, the non-uniform flows driven by asymmetric tractions ultimately cause the shape asymmetry.

Before solving \cref{eq: shape} numerically, we further illustrate the mechanism of shape asymmetry with a perturbative calculation in \cref{sec: perturbation_approach}. We expand in the small, uniform velocity difference $\delta V = \Vt - \Vavg$ between the translation velocity $\Vt = v(L) = v(-L)$, and the average basal velocity $\Vavg = \frac{1}{2L} \int_{-L}^{L} v(x)\, \dd{x} < \Vt$ (see \cref{fig: velocity_field}). By linearizing around the symmetric droplet shape, we analytically obtain the asymmetric shape to first order in $\delta V$. This calculation reveals that the location of the droplet's highest point is shifted towards the rear due to mass transport produced by the mismatch of the translation and average velocities.

\bigskip

\label{sec:solution}

\begin{figure}[tbp!]
\begin{center}
\includegraphics[width=\columnwidth]{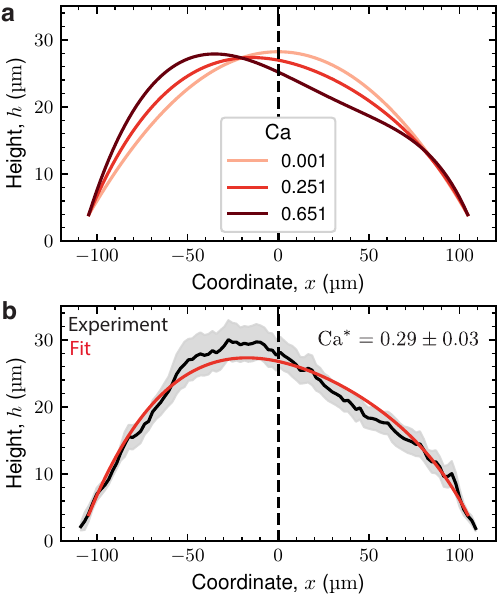}
\end{center}
 {\phantomsubcaption\label{fig: varying_numerics}}
 {\phantomsubcaption\label{fig: best_fit_numerics}}
\bfcaption{Numerical solutions to the shape equation and fit to experimental data}{ \subref*{fig: varying_numerics},
Numerical solutions of \cref{eq: shape}. As the capillary number $\Ca$ increases, the droplet shape becomes more asymmetric, with a larger difference between the front and rear contact angles. These solutions are obtained using the basal velocity profile $v(x)$ shown in \cref{fig: velocity_field} as an input, with parameter values listed in \cref{tab: parameter values}.
\subref*{fig: best_fit_numerics},
Fit of the solution of \cref{eq: shape} to the experimental averaged height profile (\cref{fig: height_profiles_avg}). 
}\label{fig: 3_numerical_shape}
\end{figure}

\noindent\textbf{Numerical solution and fit to experiments.} We now solve \cref{eq: shape} numerically and fit it to our experimental measurements. To this end, we use the basal velocity profile in \cref{fig: velocity_field}, and we must specify three conditions given that \cref{eq: shape} is a third-order differential equation. First, we impose the height at the points $\pm x_0$, with $x_0= L - \delta$, i.e., at a distance $\delta = 5.5$ $\si{\um}$ inwards from the edges, where we set $h(\pm x_0) = h_\text{c} = 4$ $\si{\um}$ according to the measurements in \cref{fig: height_profiles_avg}. We do not evaluate the height exactly at the edges to avoid dividing by zero by setting $h(L)=0$ in \cref{eq: shape}. Second, we constrain the area under $h(x)$ to correspond to the measured one from \cref{fig: height_profiles_avg}: $\int_{-x_0}^{x_0} h(x)\dd x = 4209$ $\si{\um}^2$. With these conditions, we solve \cref{eq: shape} numerically using the shooting method in Julia's differential equation package (\cref{methods}) \cite{Rackauckas2017}. As we increase the capillary number, the shape becomes increasingly asymmetric, with the highest point moving towards the rear (\cref{fig: varying_numerics}). One way to increase the capillary number $\Ca =  \mu \Vt/\gamma$ is to increase the migration speed $\Vt$. If $\Vt$ increases while keeping the velocity modulation $v(x)/\Vt$ fixed, the faster a cluster migrates, the more asymmetric its shape.

Interestingly, both the front and rear contact angles increase slightly with respect to those of the symmetric shape (\cref{fig: varying_numerics}). This observation is explained as follows: Close to the front (rear), the basal cell layer flows slightly faster (slower) than the contact line itself (\cref{fig: velocity_field}). Thus, both contact lines behave like advancing contact lines, whose dynamic contact angles are indeed larger than the static ones \cite{Leger1992}. Even though both contact angles increase with respect to the symmetric shape, the rear contact angle becomes larger than the front one (\cref{fig: varying_numerics}), consistent with our experimental observations.

Finally, we fit the numerical solution to the experimental height profile (\cref{fig: best_fit_numerics}). The theory fits the data well, and the best fit gives $\Ca^* = 0.29\pm0.03$. Since we know the average speed of the cluster, $\Vt = 50 \, \si{\um/h}$, and we set the value of surface tension to be $\gamma = 0.394  \, \si{mN/m}$ to ensure a steady cluster length (\cref{sec: parameter_estimation}), we use the fit result to estimate tissue viscosity: $\mu \approx 8.2 \, \si{kPa}\, \si{s} $. To our knowledge, no measurements of NC cell cluster viscosity have yet been reported. Previous experiments measured the viscosity of aggregates of other cell types. Based on either micropipette-aspiration \cite{Forgacs1998,Guevorkian2010,Stirbat2013,Kashkooli2021,Yousafzai2022,Boot2023,Boot2024} or compression by parallel plates \cite{Marmottant2009}, they reported viscosities in the range $\mu \approx 10^3 - 10^5 \, \si{Pa}\, \si{s} $. Our estimate is within and towards the lower end of this range, which is consistent with the fact that NC cells exhibit transient and weak cell-cell adhesion compared to standard epithelial cell types \cite{Theveneau2010,Barriga2018,Marchant2022,Kerosuo2012}.

\bigskip

\noindent\textbf{Discussion.} We reported that migrating cell clusters have asymmetric height profiles, with smaller contact angle at the front that at the rear. This asymmetric shape contrasts with the symmetric spherical-cap shape assumed to arise from surface tension in theoretical models so far. To explain the asymmetric cluster shape, we extended the theory of active wetting for tissues to let the cluster's upper surface be shaped by tissue flows instead of just surface tension. Using the physics of thin films and wetting, we obtained an equation that determines the cluster height profile, whose solution fits our experimental measurements. Our model suggests that shape asymmetry can arise simply from the asymmetric traction forces that drive cluster migration. This traction asymmetry drives forward tissue flow, which creates a pressure gradient across the tissue. Through the Young-Laplace equation, this pressure gradient shapes the cluster's upper surface, which becomes more curved at the rear than at the front, thus resulting in the observed shape asymmetry.

Whereas our experimental observations focus on durotaxis, the mechanism that we propose theoretically is generic, as it relies on the principle of migration by asymmetric tractions \cite{Rossetti2024}. Collective chemotaxis, haptotaxis, and electrotaxis all involve symmetry breaking \cite{Espina2022}, and they could therefore also rely on asymmetric tractions. Thus, we speculate that our findings could apply to any droplet-like migrating cell cluster, regardless of the stimulus that guides it.

Our work uncovers a view of migrating cell clusters as active droplets moving on a wettability gradient. The guidance cue --- be it chemoattractant, substrate stiffness, electrostatic potential, or others --- leads to stronger active traction at the tissue front, which locally increases the tissue wettability \cite{Perez-Gonzalez2019,Franco-Onate2026}. Effectively, the tissue moves thanks to this active wettability gradient. In this sense, our results provide an active-matter analogue to the motion of passive droplets on wettability gradients created via temperature gradients, surface coatings, or chemical reactions that locally modify surface tension \cite{deGennes2004,Greenspan1978,Raphael1988,Brochard1989,Ondarcuhu1991,Chaudhury1992,DosSantos1995,Thiele2004,Lazar2005,Subramanian2005,Pismen2006,Moumen2006}. Unlike in passive droplets, however, the motion of our active droplets results from global force balance across the entire tissue rather than local balance of surface tensions at the contact line. Moreover, the no-slip boundary condition used to model the adhesion of passive liquids on substrates is replaced here by the migration velocity field of the basal cells on the substrate. These changes, along with the polar active forces that drive cell migration, allowed us to extend the classical results to describing tissues.

Our work captures the interplay between active forces, flows, and shape in tissues. This interplay has also been studied in other models of active droplets \cite{Joanny2012,Khoromskaia2015,Trinschek2017,Trinschek2020,Loisy2019,Loisy2020,Shankar2022,Stegemerten2022,Coelho2023,Ziethen2024,Li2025,Ford2025,Liese2025,Martinez-Calvo2026,Panigrahi2025,Celora2026}, which have generally considered active driving throughout the droplet rather than active traction at its base. Moreover, except for recent work on collective chemotaxis \cite{Ford2025,Panigrahi2025,Celora2026}, active droplet models have focused on how self-propulsion arises from spontaneous symmetry breaking \cite{Trinschek2020,Loisy2019,Loisy2020,Stegemerten2022} rather than from a prescribed asymmetry as done here for guided collective cell migration.

Future theoretical work could generalize our model by studying the shear stresses exerted by the upper cell layers on the basal one, thus completing the mechanical feedback between cell layers. Other possible directions include to analyze the cluster's in-plane shape \cite{Sanoria2026,Termens2025}, and to account for elastocapillary effects resulting from cell-generated substrate deformations \cite{Yousafzai2022a,Shi2022,Chandel2025}. Already for passive droplets, gradients of substrate deformation can result in durotactic motion \cite{Style2013,Bueno2018,Gomez2020,Zhao2025b}. In cell aggregates, coupling asymmetric active tractions, tissue shape, and substrate deformation is therefore an interesting direction for future work. Finally, on the experimental front, our work poses the challenge to measure cell flows across the tissue height to compare them with the theoretical predictions. From a broader perspective, pursuing these directions will advance our understanding of migrating cell collectives as active droplets that shape themselves.

\bigskip


\noindent\textbf{Acknowledgments.}
We thank Pierre Haas and Jean-Fran\c{c}ois Joanny for discussions.
R.A. acknowledges funding from the European Union through the ERC Starting Grant `Living Fluctuations'.
This work was performed in part at Aspen Center for Physics, which is supported by National Science Foundation grant PHY-2210452.

\bigskip

\noindent\textbf{Author contributions.} J.A.E. and E.H.B. performed the experiments. H.I.H. and R.A. developed the model and fitted it to the experimental data. H.I.H. and R.A. wrote the paper with input from J.A.E. and E.H.B. All authors jointly conceived the project and discussed the results.

\bibliography{shape_asymmetry_abbrev,Preprints,shape_asymmetry}

\onecolumngrid
\clearpage

\appendix

\setcounter{equation}{0}
\setcounter{figure}{0}
\renewcommand{\thefigure}{S\arabic{figure}}

\onecolumngrid
\begin{center}
\textbf{\large Supplementary Material}
\end{center}

\twocolumngrid

\section{\MakeUppercase{Methods}} \label{methods}

\subsection{Durotactic gels}

Gels with stiffness gradients were prepared from two polyacrylamide solutions. For soft gels: $550\, \si{\mu l}$ of $7.6\, \si{mM}$ hydrochloric acid (HCl), $319.5\, \si{\mu l}$ H$_2$O, $0.5\, \si{\mu l}$ tetramethylethylenediamine (TEMED), $75\, \si{\mu l}$ of $40\%$ acrylamide, $30\, \si{\mu l}$ $2\%$ bis-acrylamide, $20\, \si{\mu l}$ $0.1\, \si{M}$ N-hydroxysuccinimide (NHS). For stiff gels: $550\, \si{\mu l}$ of $7.6\, \si{mM}$ HCl, $258\, \si{\mu l}$ H$_2$O, $0.5\, \si{\mu l}$ TEMED, $137.5\, \si{\mu l}$ of $40\%$ acrylamide $25\, \si{\mu l}$ $2\%$ bis-acrylamide, $20\, \si{\mu l}$ of  $0.1\, \si{M}$ NHS. Then $4\, \si{\mu l}$ of $200\, \si{nm}$ carboxylated-modified acrylic beads were added for correct gradient formation of the gels. Gels were polymerized by adding $5\, \si{\mu l}$ of $10\%$ ammonium persulfate. The still liquid mixes were added sequentially to a mold as previously described \cite{Shellard2021}, where polymerization proceeded for $45\, \si{min}$ at room temperature in a humidifier chamber. 

Fibronectin was covalently linked by immersion of the gels into a solution containing $0.2\, \si{M}$ EDC, $0.1\, \si{M}$ NHS and $0.1\, \si{M}$ 2-morpholinoethanesulfonic acid hydrate (MES) buffer. After washing twice with Phosphate-Buffered Saline (PBS), gels were incubated with $0.1\, \si{mg} \, \si{ml}^{-1}$ of fibronectin for $1\, \si{h}$ $45\, \si{min}$ at room temperature. Fibronectin was washed with PBS and the crosslinking reaction was quenched by incubating the gels for $15\, \si{min}$ with $0.32\%$ ethano-lamine in PBS. Gel functionalization and effective Young's modulus were checked as previously described \cite{Shellard2021, Marchant2022}.

\subsection{Embryo obtention and NC disssection}

Adult animals were maintained at $18\degree  \si{C}$ in a temperature-controlled environment and embryos were obtained by in vitro fertilization. Briefly, induction of ovulation was performed by injecting chorionic gonadotrophin into adult females; after ovulation, oocytes were in vitro fertilized by mixing with a sperm solution. On successful fertilization, embryos were injected with nuclear and membrane fluorescent markers and staged by following developmental tables \cite{alma995875444202441} while maintained between $14$ and $18 \degree  \si{C}$.

The neural crest was dissected as previously described \cite{Barriga2018}. Briefly, using a hair knife, embryo epidermis was lifted and the NC underneath was removed.

\subsection{Fixing of cluster and imaging}

Dissected explants were transferred to a dish containing the gel in Danilchik's For Amy (DFA) medium. Here we waited $4\si{h}$ to allow the explant to attach and to directionally migrate along the stiffness gradient. Once migrating we fixed the cells as previously described \cite{Barriga2018}. Fixed explants were treated with DAPI (4′,6-diamidino-2-phenylindole) and imaged from top to bottom to acquire the full height of the cluster in a Leica Stellaris 5 confocal system by using a 20x water immersion objective. Height profiles from z-stack images were obtained using a combination of built-in imageJ and MATLAB-based plugins. 

\subsection{Contact angle and peak position measurements}

\begin{figure*}[tbp!]
\begin{center}
\includegraphics[width=\textwidth]{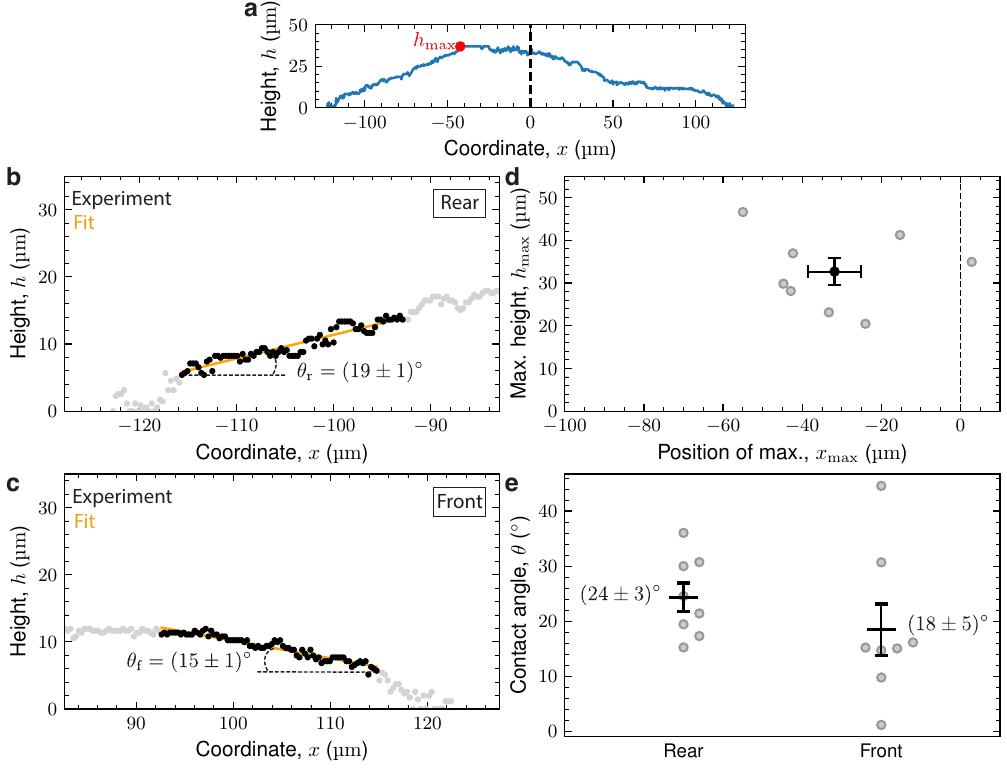}
\end{center}
 {\phantomsubcaption\label{fig: height_individual}}
 {\phantomsubcaption\label{fig: theta_r_individual}}
 {\phantomsubcaption\label{fig: theta_f_individual}}
 {\phantomsubcaption\label{fig: hmax_xmax}}
  {\phantomsubcaption\label{fig: thetaf_theta_r}}
\bfcaption{Contact angles and peak position for individual clusters}{ \subref*{fig: height_individual}, Height profile of a representative cell cluster. The position of the peak is marked with a red point. 
\subref*{fig: theta_r_individual},\subref*{fig: theta_f_individual}, Contact angle fits in the rear and front. The gray points are excluded from the fit (see \cref{methods}).
\subref*{fig: hmax_xmax}, Position of the highest point for each cluster (gray). The black point shows the mean and standard error of the mean.
\subref*{fig: thetaf_theta_r}, Rear and front contact angles for each cluster (gray). The black points show the mean and standard error of the mean.}
\label{fig: individual contact angles}
\end{figure*}

\begin{figure}[tbp!]
\begin{center}
\includegraphics[width=\columnwidth]{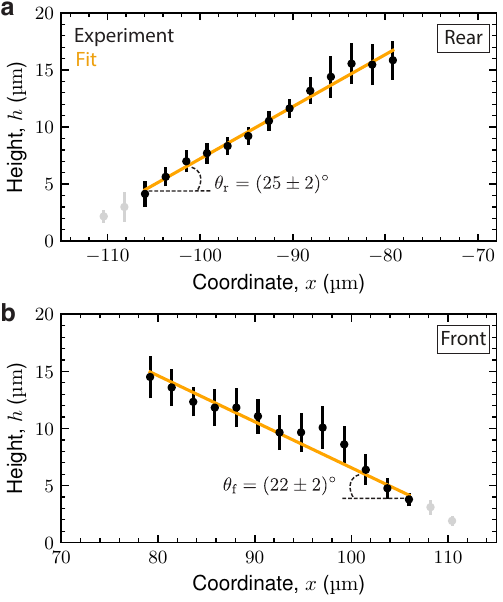}
\end{center}
 {\phantomsubcaption\label{fig: rear_contact_angle}}
 {\phantomsubcaption\label{fig: front_contact_angle}}
\bfcaption{Rear (\subref*{fig: rear_contact_angle}) and front (\subref*{fig: front_contact_angle}) contact angles from the averaged height profile}{ The black points show the mean and the standard error of the mean. The gray points are excluded (see \cref{methods}).}
\label{fig: contact angles}
\end{figure}

We measure the front and rear contact angles by fitting a straight line to the height profile close to the edges (\cref{fig: individual contact angles,fig: contact angles}). We take the data points up to $30\, \si{\um}$ away from the edge. We also exclude an edge region below a threshold height of $4.5\, \si{\um}$ (gray data points in \cref{fig: individual contact angles,fig: contact angles}) to avoid capturing the detailed shape of the edge cell rather than the cluster's overall shape. We measured the contact angles both for individual clusters (\cref{fig: individual contact angles}) and for the length-normalized and averaged height profile shown in \cref{fig: height_profiles_avg} (\cref{fig: contact angles}).

\subsection{Numerical solver}

We use a shooting method to numerically solve \cref{eq: shape}. We implement the problem using the \texttt{Shooting} solver of \texttt{BoundaryValueDiffEq.jl}~\cite{Rackauckas2017} with the underlying initial-value integration performed by the explicit Bogacki--Shampine third-order Runge--Kutta scheme (BS3).
Starting from $h(x_\text{r}) = h_\text{c}$, the shooting parameters are the slope  $h'(x_\text{r})$ and curvature $h''(x_\text{r})$, which are iteratively adjusted until the solution satisfies $h(x_\text{f}) = h_\text{c}$ and $\int_{x_\text{r}}^{x_\text{f}} h(x)\,\mathrm{d}x = A$, with $x_\text{r} = 0 $ and $x_\text{f} = 2L$.

\section{\MakeUppercase{Analytical solution for the basal velocity field}} \label{sec: velocity_solution}

Here, we solve the force balance \cref{eq: 1d force balance}. Together with the expression of the viscous tension $\sigma=\eta \partial_x v$, it reads
\begin{equation}
\eta \partial_x^2 v(x) = \xi v(x) - \Ta(x) p(x).
\end{equation}
As specified in the Main Text, we take a polarity field $p(x) = x/L$ and an active traction magnitude given by \cref{eq: traction_front_back}. The solution reads
\begin{equation}
v(x) =
\begin{cases}
C_1 \sinh{(\frac{x}{\lambda})} + C_2 \cosh{(\frac{x}{\lambda})} + \frac{T_\text{r}}{\xi}\frac{x}{L},  & x \in [-L,0] \\
C_3 \sinh{(\frac{x}{\lambda})} + C_4 \cosh{(\frac{x}{\lambda})} + \frac{T_\text{f}}{\xi}\frac{x}{L}, & x \in [0,L],
\end{cases}
\end{equation}
where $C_1,\ldots, C_4$ are integration constants. They are determined by imposing the boundary conditions \cref{eq Young-Dupre} as well as continuity conditions for both the stress and velocity fields across the boundary between high and low tractions at $x=0$:
\begin{align}
    v_{\text{r}}(0) &= v_{\text{f}}(0), \\
    \sigma_{\text{r}}(0) &= \sigma_{\text{f}}(0),
\end{align}
where the subscripts refer to the rear and front parts of the solution. We obtain
\begin{subequations}
\begin{align}
    C_1 = - \frac{\lambda}{\cosh{(\frac{L}{\lambda})}} \left( \frac{\gamma}{2 \eta} (\cos{\theta_\text{f}} + \cos{\theta_\text{r}}) + \frac{ \overline{T}}{\xi L }\right) + \frac{\lambda \, \tilde{T}}{ \xi L},\\
    C_3 =  - \frac{\lambda}{\cosh{(\frac{L}{\lambda})}} \left( \frac{\gamma}{2 \eta} (\cos{\theta_\text{f}} + \cos{\theta_\text{r}}) + \frac{ \overline{T}}{\xi L }\right) - \frac{\lambda \, \tilde{T}}{ \xi L},\\
    \begin{multlined}
    C_2 = C_4 =  - \frac{\lambda}{\sinh{(\frac{L}{\lambda})}} \left( \frac{\gamma}{2 \eta} (\cos{\theta_\text{f}} - \cos{\theta_\text{r}}) + \frac{ \overline{T}}{\xi L }\right) \\
      + \frac{\lambda \, \tilde{T}}{ \xi L}\frac{1}{ \tanh{(\frac{L}{\lambda})}}.
     \end{multlined}
\end{align}
\end{subequations}
Here, we have introduced the average and difference of the front and back traction $\overline{T}  = \frac{T_\text{f} + T_\text{r}}{2}$ and $\tilde{T}= \frac{T_\text{f} - T_\text{r}}{2}$, as well as the hydrodynamic screening length $\lambda = \sqrt{\eta/\xi}$.

From the solution for the velocity field, we obtain the spreading velocities $\Vs$:
\begin{multline}\label{eq: spreading_V}
    \Vs = \frac{v(L) - v(-L)}{2} = \\
    \frac{\overline{T}}{\xi} \left(1- \frac{\lambda}{L} \tanh\!{\left(\frac{L}{\lambda}\right)} \right)
    -  \frac{\gamma}{\xi \lambda}\frac{(\cos{\theta_\text{f}} + \cos{\theta_\text{r}})}{2}\tanh\!{\left(\frac{L}{\lambda}\right)}.
\end{multline}
When the spreading velocity vanishes, $\Vs=0$, the cluster reaches a steady-state length. In that case, the translation velocity $\Vt$ is given by
\begin{multline}\label{eq: translation_V}
    \Vt = v(L) = v(-L) = \\
    \frac{\tilde{T} }{\xi} \left(1 - \frac{\lambda}{L} \tanh{\left(\frac{L}{2\lambda}\right)}\right)
    -  \frac{\gamma}{\xi \lambda}\frac{(\cos{\theta_\text{f}} - \cos{\theta_\text{r}})}{2} \coth{(\frac{L}{\lambda})}.
\end{multline}
The translation velocity depends on the average traction and the sum of the surface tension contributions, whereas the spreading velocity depends on the difference of both tractions and surface tension contributions from the front and rear. We also point out that the translation velocity, i.e. the velocity of the front and rear edges, is always larger than the average velocity $\Vavg$, which is given by
\begin{equation} \label{eq: avg_V}
    \Vavg =\frac{1}{2L} \int_{-L}^L v(x) \dd x = \frac{1}{2L \, \xi} \left(\tilde{T} L - \gamma(\cos{\theta_\text{f}} - \cos{\theta_\text{r}}) \right).
\end{equation}

\section{\MakeUppercase{Parameter estimation}} \label{sec: parameter_estimation}
Here, we estimate the parameter values of our model to match experimental data. 
Note that the mechanism to generate a shape asymmetry is general and does not rely on a specific choice of parameters; it follows from any non-uniform translation velocity of the basal monolayer, which our model predicts to arise from a front-back asymmetry in active traction forces.

\begin{figure}[tbp!]
\begin{center}
\includegraphics[width=\columnwidth]{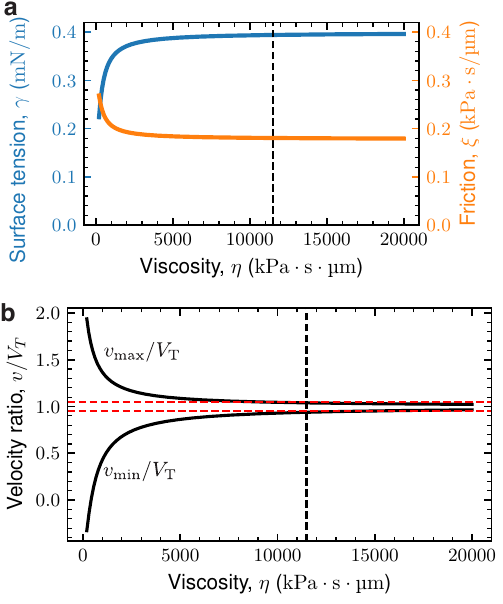}
\end{center}
 {\phantomsubcaption\label{fig: surface_tension_friction_vs_eta}}
 {\phantomsubcaption\label{fig: velocity_ratio_eta}}
\bfcaption{Setting monolayer viscosity}{ \subref*{fig: surface_tension_friction_vs_eta}, Values of surface tension $\gamma$ and friction coefficient $\xi$ required to satisfy our requirements $\Vs=0\, \si{\um/h}$ and $\Vt = 50\, \si{\um/h}$ for the spreading and translation velocity, for any chosen value of the monolayer viscosity $\eta$. \subref{fig: velocity_ratio_eta}, Ratio of the maximal and minimal velocity along the profile in \cref{fig: velocity_field} with respect to the translation velocity $\Vt$, for any chosen value of the monolayer viscosity $\eta$. Dashed red lines indicate $\pm 5\%$ of $\Vt$. The vertical dashed line indicates the value of viscosity $\eta$, and hence of $\xi$ and $\gamma$, chosen for \cref{fig: velocity_field} in the Main Text.}
\label{fig: parameter_estimates}
\end{figure}

\begin{table}[tbp!]
    \centering
    \begin{tabular}{l|l|l}
         \textbf{Symbol} & \textbf{Description} & \textbf{Estimate} \\
         \hline
         $L$ & Cluster half-length& $110 \,  \si{\um}$\\
         $T_\text{f}$ & Active traction front half& $15 \,  \si{Pa}$ \\
         $T_\text{r}$& Active traction rear half& $5 \,  \si{Pa}$\\
         $\eta$ & Monolayer viscosity&  $ 11487 \, \si{kPa}\, \si{s} \, \si{\um}$\\
         $\xi$ & Friction coefficient&  $0.18 \, \si{kPa} \, \si{s/\um}$ \\
         $\lambda$ & Hydrodynamic screening length& $252.6 \,  \si{\um}$\\
         $\gamma$ & Surface tension&  $0.394 \,  \si{mN/m}$ \\
         $\mu$ & Cluster viscosity&  $ 8.2 \, \si{kPa}\, \si{s}$\\
    \end{tabular}
    \bfcaption{Parameters estimates}{ The choice of these values is discussed in the text.}
    \label{tab: parameter values}
\end{table}

We first estimate the front and back active traction magnitudes to be $T_\text{f} = 15\,\si{Pa}$ and $T_\text{r} = 5\,\si{Pa}$ from hyper- and hypo-acetylated NC cell clusters, which arise in stiff and soft substrates, respectively \cite{Marchant2022}. We set the tissue half-length to be $L=110\,\si{\um}$ as measured in our experiments. We also take the rear and front contact angles measured from the averaged height profile, which gives $\theta_\text{r} = 25\degree$ and $\theta_\text{f} = 22\degree$ (\cref{fig: contact angles}).
Note that these values are used here only to parametrize the basal velocity field $v(x)$, and not as boundary conditions for the cluster shape $h(x)$. We do not have direct measurements of the remaining model parameters. In the following, we estimate them by requiring that (i) the tissue reaches a steady-state length, implying that the spreading velocity vanishes: $\Vs=0$, and (ii) that the cell cluster migrates at the measured speed of $\Vt \approx 50\, \si{\um/h}$.

Imposing these two requirements through \cref{eq: spreading_V,eq: translation_V}, and given that we know $T_\text{f}$, $T_\text{r}$, $L$, $\theta_\text{f}$ and $\theta_\text{r}$, we are left with three unknown parameters: the friction coefficient $\xi$, the monolayer viscosity $\eta$, and surface tension $\gamma$. Note that \cref{eq: spreading_V,eq: translation_V} are written in terms of $\gamma$, $\xi$, and the hydrodynamic screening length $\lambda = \sqrt{\eta/\xi}$. Since we have two equations and three unknowns, we have freedom to choose the value of one of the parameters. To address this issue, we vary tissue viscosity and study how the values of surface tension $\gamma$ and friction coefficient $\xi$ that result from imposing the two requirements above (\cref{fig: parameter_estimates}). \Cref{fig: surface_tension_friction_vs_eta} shows that, for sufficiently high viscosity, the values of $\xi$ and $\gamma$ saturate, implying that their values are robust to changes in viscosity in this regime. Viscosity also tunes the amplitude of the velocity profile in \cref{fig: velocity_field}. We plot the maximal and minimal velocity along the profile in \cref{fig: velocity_ratio_eta}. Since our experiments show very little velocity variations within the cluster, we choose a value of tissue viscosity so that the amplitude of the velocity profile is $5\%$ of the migration velocity $\Vt$, as indicated by the dashed red lines in \cref{fig: velocity_ratio_eta}. The chosen parameter values are collected in \cref{tab: parameter values}.

Note that choosing a different amplitude of the velocity profile modulation would lead to a different viscosity estimate, which can be seen from \cref{fig: velocity_ratio_eta}. This estimate is therefore not robust. However, for small velocity modulations, varying monolayer viscosity would entail only a small change in surface tension $\gamma$ and friction coefficient $\xi$, which are robust since they are relatively insensitive to viscosity variations for sufficiently large viscosities (\cref{fig: surface_tension_friction_vs_eta}). Thus, the monolayer viscosity $\eta$ essentially controls the monolayer velocity modulation, which remains uncertain, whereas $\gamma$ and $\xi$ robustly ensure that the cluster achieves a steady-state length and migration speed. The mechanism for shape asymmetry that we reveal, however, is generic and does not rely on specific parameter values.

\section{\MakeUppercase{Asymmetric shape by perturbing around the symmetric shape}} \label{sec: perturbation_approach}

To further illustrate the mechanism of shape asymmetry, here we obtain asymmetric shapes by perturbing the shape equation (\cref{eq: shape}) around the symmetric shape $h_0(x)$, for which $h_0'''(x)=0$. Solving this equation, the symmetric shape is given by $h_0(x) = a x^2 + bx + c$. Here, $a,b,c$ are integration constants, which we determine by imposing the boundary conditions $h_0(\pm x_0) = h_\text{c}$ and the area condition $\int_{-x_0}^{x_0} h_0(x) \dd x = A$, as described in the Main Text. Thus, the solution for the symmetric equation reads
\begin{equation} \label{eq: shape_h0}
    h_0(x) = -\frac{3 (A - 2 \hc x_0)}{4 x_0^3} \left(x^2 -x_0^2 \right) + \hc,
\end{equation}
which can be rewritten in terms of its roots as
\begin{multline}\label{eq: shape_h0_with_radicals}
    h_0(x) = \frac{3 (A - 2 \hc x_0)}{4 x_0^3} \left(x_0^2 + \frac{4 x_0^3}{3 (A - 2 \hc x_0)} \hc -x^2\right) \\
    = q (r-x)(r+x)
\end{multline}
with $r= \sqrt{x_0^2 + \frac{4 x_0^3}{3 (A - 2 \hc x_0)} \hc}$ and $q = \frac{3 (A - 2 \hc x_0)}{4 x_0^3}$.

\begin{figure}[tbp!]
\begin{center}
\includegraphics[width=\columnwidth]{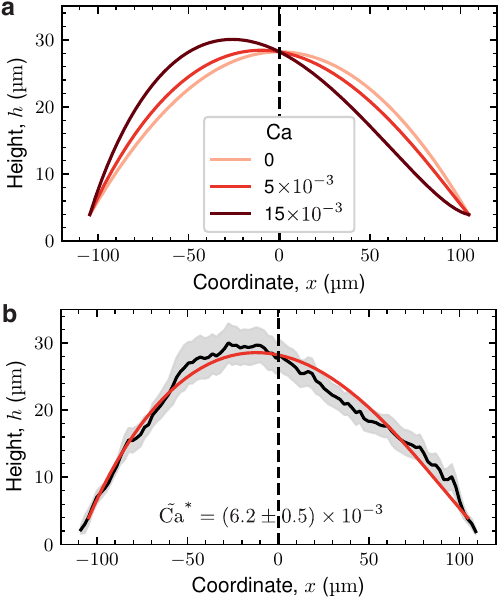}
\end{center}
 {\phantomsubcaption\label{fig: perturbation_varying_Ca}}
 {\phantomsubcaption\label{fig: perturbation_fit}}
\bfcaption{Perturbative solution for the cluster shape}{ \subref*{fig: perturbation_varying_Ca}, By increasing the capillary number, the height profile becomes asymmetric. 
\subref*{fig: perturbation_fit}, Fit of the perturbative solution \cref{eq: h_perturbation_full} to the experimental averaged height profile.} \label{fig: perturbation}
\end{figure}

We next look for perturbations around this symmetric solution. \Cref{eq: shape} shows that asymmetric shapes arise from differences between the basal layer velocity $v(x)$ and the migration velocity $\Vt$. For the sake of simplicity, we consider a uniform and small velocity difference $\delta V = \Vt - \Vavg$ (\cref{fig: velocity_field}), which we use as a perturbation. Linearizing \cref{eq: shape}, the corresponding height perturbation $\delta h(x)$ satisfies
\begin{equation} \label{eq: height_perturbation}
\delta h'''(x) = 3 \Ca \frac{\delta V/\Vt}{h_0^2(x)}.
\end{equation}
Introducing \cref{eq: shape_h0_with_radicals}, we solve this equation by integrating it three times using
\begin{subequations}
\begin{align}
&\int \frac{\dd x}{zt} = \frac{1}{\Delta} \ln \frac{t}{z},\\
&\begin{multlined}
\int \frac{\dd x}{z^n t^m} = -\frac{1}{(m - 1)\Delta} \cdot \frac{1}{t^{m-1} z^{n-1}} \\
- \frac{(m + n - 2)b}{(m - 1)\Delta} \int \frac{\dd x}{t^{m-1} z^{n}},
\end{multlined}
\\ 
&\begin{multlined}
\int \frac{\dd x}{t^{m-1} z^n} = \frac{1}{(n - 1)\Delta}\frac{1}{t^{m-1} z^{n-1}} \\
+ \frac{(m + n - 2)\beta}{(n - 1)\Delta} \int \frac{dx}{t^{m} z^{n-1}},
\end{multlined}
\end{align}
\end{subequations}
where $z = a+b x$, $t = \alpha + \beta x$, $\Delta = a \beta - \alpha b$, and $m=n=2$ (relations 2.154 and 2.155 in Ref. \cite{Gradshteyn2007}). We thus find the solution
\begin{multline}
    \delta h(x) = 3 \Ca \frac{\delta V}{\Vt} \frac{1}{4 r^3 q^2} \left( \frac{1}{2} (r^2 - x^2) \left[ \ln(r-x) \right.\right. \\
    \left. \left. - \ln(r+x)) \right] + \frac{c_1}{2} x^2 + c_2 x + c_3 \right).
\end{multline}
The boundary conditions for the perturbation $\delta h$ are chosen such that the total height profile $h(x) = h_0(x) + \delta h(x)$ fulfils the boundary conditions described in the Main Text, namely $h(\pm x_0) = h_\text{c}$ and $\int_{-x_0}^{x_0} h(x)\dd x = A$. Given that we already imposed such boundary conditions on the symmetric, unperturbed solution $h_0(x)$ in \cref{eq: shape_h0} above, for $\delta h(x)$ we have
\begin{subequations}
\begin{align}
    \delta h(x = -x_0) =0, \\
    \delta h(x = x_0) =0, \\
    \int_{-x_0}^{x_0} \delta h(x) \, \dd x = 0.
\end{align}
\end{subequations}
Imposing these conditions, the integration constants are given by
\begin{subequations}
\begin{align}
&    c_1 = - \frac{6 c_3}{x_0^2} \label{eq: pert_c1}, \\
&\begin{multlined}
    c_2 = \frac{1}{2 x_0} \left\{ (r^2 -x_0^2) \left[ \ln(r+x_0) \right. \right. \\
      \left. \left. - \ln(r-x_0) \right] + x_0^2 c_1 + 2 c_3\right\}, \label{eq: pert_c2}
\end{multlined}\\
&    c_3 = -c_1 \frac{x_0^2}{2}. \label{eq: pert_c3}
\end{align}
\end{subequations}
From \cref{eq: pert_c1,eq: pert_c3}, we conclude $c_1 = c_3 = 0$.
Therefore, the solution for the shape perturbation is
\begin{multline}
    \delta h(x) = 3 \Ca\frac{\delta V}{\Vt} \frac{1}{8 r^3 q^2} \left\{ \left(r^2 - x^2\right) \ln\left(\frac{r-x}{r+x}\right) \right. \\
    \left.+ \frac{x}{x_0} \left[ \left(r^2 -x_0^2\right) \ln\left(\frac{r+x_0}{r-x_0}\right) \right] \right\}.
\end{multline}
Finally, the total perturbed asymmetric shape is
\begin{multline} \label{eq: h_perturbation_full}
    h(x) = q(r-x)(r+x) \\
    + 3 \Ca\frac{\delta V}{\Vt} \frac{1}{8 r^3 q^2} \left\{ \left(r^2 - x^2\right) \ln\left(\frac{r-x}{r+x}\right) \right. \\
    \left. + \frac{x}{x_0} \left[ \left(r^2 -x_0^2\right) \ln\left(\frac{r+x_0}{r-x_0}\right) \right] \right\}.
\end{multline}
We plot this height profile in \cref{fig: perturbation_varying_Ca}, which shows how the shape asymmetry arises as we increase the perturbation capillary number $\tilde{\Ca} = \Ca\, \delta V/\Vt$. Therefore, as the difference between the translation and average velocities slowly increases, the asymmetry becomes larger. 
The highest point is shifted rearwards if $\Vt > \Vavg$, and the asymmetry would be flipped if $\Vt < \Vavg$.
In particular, for $\Vt>\Vavg$, the contact angle at the front decreases with respect to the unperturbed shape, whereas the contact angle at the rear increases.

This perturbative approach is useful in showing that the difference between the flow in the basal layer and the translation velocity gives rise to the observed shape asymmetry. 
It also highlights that the rearward shift of the highest point is a consequence of $\Vt > \Vavg$. 
However, the perturbative treatment is limited to small perturbations $\delta V = \Vt - \Vavg$. To go beyond it, in the Main Text, we present numerical solutions to the full nonlinear \cref{eq: shape}.

Finally, we fit \cref{eq: h_perturbation_full} to the averaged experimental height profile data (\Cref{fig: perturbation_fit}).
We obtain a best fit for $\tilde{\Ca}^* = (6.2\pm 0.5)\times 10^{-3}$, which is three orders of magnitude lower compared to value of $\Ca$ obtained in \cref{fig: best_fit_numerics}.
This is a consequence of the definition of $\tilde{\Ca} = \Ca\, \delta V/\Vt$, where  $\delta V/\Vt\ll 1$.

\onecolumngrid

\clearpage

\onecolumngrid
\begin{center}
\textbf{\large Supplementary Videos} \label{movies}
\end{center}



\noindent\textbf{Video 1. Three-dimensional shape of a migrating cell cluster.} Three-dimensional confocal microscopy image of a neural crest cell cluster fixed while it was performing durotaxis. Cell nuclei are fluorescently labeled in blue (DAPI), and cell membranes in green (membrane-GFP).

\end{document}